# Entropy variation in a fractal phase space

Oscar Sotolongo-Costa[1], Isaac Rodríguez Vargas[2]

*1..-Cátedra "Henri Poincaré" de sistemas complejos. Havana University. Habana 10400. Cuba..*
osotolongo@gmail.com

*2.- Unidad Académica de Ciencia y Tecnología de la Luz y la Materia, Universidad Autónoma de Zacatecas, 98160, Zacatecas, Zac., Mexico.*isaac@uaz.edu.mx

## Abstract

In this work, with the help of fractional calculus, it is shown a time dependence of entropy more general than the well known Pesin relation is derived. Here the equiprobability postulate is not assumed, the system dynamic in the phase space is not necessarily Markovian and the system is not in a steady state at all. Different possibilities for the time evolution of entropy by considering different features of the phase space and processes involved are obtained.

## 1 Introduction

Entropy is one of the most fascinating, abstract and complex concepts in physics. Among the most important characteristics of entropy its extensive and non-conserved character stand out. From a microscopic stand point entropy can be linked to the probabilistic features of the accessible microstates of a system, *i.e.,* to the peculiarities of the corresponding phase space.

In nonlinear dynamics the evolution of entropy is a linear function of time or equivalently the entropy production rate is constant, known as Kolmogorov-Sinai entropy [1, 2].

The aim of this work is to briefly show that other rates of entropy variation with time are possible when the geometry of the phase space is fractal.

## 2 The model

The time evolution of the entropy of an arbitrary system is analyzed by studying the system evolution towards equilibrium, through the diffusion of the representative points in the phase space. The possible set of initial conditions of the system are represented by a set of phase space points. The time evolution of these points takes place diffusively.

We must notice that fine-grained quantities do not vary with time, since, by Liouville's Theorem, the volume of phase space occupied by the system during its evolution is constant. Instead, under coarse-graining, *i.e.*, smearing or smoothing of the probability distribution in phase space, the occupied volume keeps increasing as discussed in [3,10]

The simplest way to produce a coarse-graining is to divide the phase space in cells such that the sum of their volumes equals the total volume of the available phase space [3].

In a coarse-grained phase space the smearing out of the initial points leads to an increase of volume containing these points, so the entropy as a measure of this volume shall vary with time. We may omit the markovian evolution of the diffusion accepting that the system has some memory of its evolution, so the number of accessible points in phase space vary as

$$N(t) - N_0 = \int dx' \int_0^t F(x,x',t,u)N(u)du, \quad (1)$$

$$F(x,t,x',u) = \frac{g(x')}{\Gamma(\nu)}(t-u)^{\nu-1}, \quad (2)$$

where $N(t), N_0$ are the number of accesible points at $t$ and at the initial time, and $F(x,t,x',u)$ is a propagator that involves the points of the phase space ($x$) and time ($t$).

$g(x)$ is the density of states and $\Gamma(\nu)$ is the Euler Gamma function. Note that we are repeating the symbol $\Gamma$ for the volume in phase space and for gamma-function. However, this is made in the appropriate context each moment. Hope there is no confusion.

Substitution of (2) in (1) gives a standard Riemann fractional integral (See [5]). This can be solved by Laplace transformation, leading to the solution [6]:

$$\begin{cases} N(t) = N_0 E_\nu(-(ct)^\nu) \\ N_0 = N(0), \end{cases} \quad (3)$$

where

$$E_\nu(x) = \sum_{k=0}^{\infty} \frac{x^{\nu k}}{\Gamma(\nu k + 1)}, \quad (4)$$

is the well-known Mittag-Leffler function, $c$ is the result of the integration on $x'$, which is assumed finite.

Is easy to see that the Mittag-Leffler function is a generalization of the exponential function. When $\nu = 1$, (4) becomes the series development of $e^x$.

Here, the time dependence of the phase space volume can be computed readily by assuming that the number of states and volume are connected by a power law $N(t) \sim \Gamma^d$. Explicitly, the variation of the accessible phase space volume becomes

$$\Gamma(t) = \Gamma_0 E_\nu^{\frac{1}{d}}\left[(ct)^\nu\right]. \quad (5)$$

Finally, it is well known that when the number of states do not grow linearly with the volume $\Gamma$, the entropy adopts the mathematical expression [7]:

$$S_d(t) = \frac{\Gamma(t)^{1-d} - 1}{1-d}. \quad (6)$$

What expresses the Tsallis entropy with index $d$ instead of $q$. (For Tsallis entropy, see [8]). Therefore,

$$S_d(t) = \Gamma_0^{1-d} l_d \left\{ E_\nu\left[\frac{(ct)^{\nu k}}{\Gamma(\nu k + 1)}\right]\right\}^{\frac{1}{d}} \quad (7)$$

(7) expresses the time variation of entropy characteristic of fractal phase space. As explained in [8], this entropy is useful to deal with a lot of problems involving chaotic systems and long range correlations in time and space. The non linear behavior of entropy found in [9], for early stages of the system, can be obtained. Here $l_d\{x\} = \dfrac{x^{1-d}-1}{1-d}$, we prefer this notation for the generalized logarithm, commonly expressed in the uncomfortable way as $\ln_q\{x\} = \dfrac{x^{1-q}-1}{1-q}$, for this last form has always been used to denote the logarithm base $q$.

This equation may be reduced to particular cases, specially when $d=1$ (short range space correlations), $\nu=1$ (no memory effects), and when both things happen the currently known Kolmogorov–Sinai entropy is obtained and the current linear dependence with time is obtained.

## 4 Conclusions

Time variation of entropy has been computationally studied in different maps [3,9,10]. Here an analytic approach based in the fractal properties of the phase space was presented.

A close expression for the time dependent entropy was obtained. This expression is derived by assuming that the number of states in the phase space undergoes a diffusion process in the fractal phase space. For brevity, we have not included the possibility of a "fractal" dynamics where the temporal dependence scales as a power of time.

## References

1. A. N. Kolmogorov, Dokl. Akad. Nauk SSSR **119**, 861–864 (1958).
2. A. N. Kolmogorov, Dokl. Akad. Nauk SSSR **124**, 754–755 (1959).
3. F. Baldovin and A. Robledo, Phys. Rev. E **69**, 045202(R) (2004).
4. A.E.M. El-Misiery, E. Ahmed, Applied Math. And Computation **178** 207-211 (2006).
5. K.B. Oldham, J. Spanier, "The Fractional Calculus" Dover Publications N.Y. (2004).
6. R.K. Saxena. A.M. Mathai, H.J. Haubold , arXiv:math-ph/0406047
7. V. Garcia-Morales, J. Pellicer, ArXiv: cond-mat/0508606v1 25 Aug 2005.
8. C. Tsallis, *"Introduction to Nonextensive Statistical Mechanics"*, Springer, New York (2009).
9. M. Baranger, V.Latora, and A. Rapisarda  Chaos, Solitons and Fractals **13**, 471-478 (2002).
10. V. Latora, M. Baranger, A. Rapisarda, C. Tsallis,  Phys. Lett. A **273**(1–2), 97–103 (2000)